# Epitaxial α-Ta (110) film on a-plane sapphire substrate for superconducting qubits on wafer scale

Boyi Zhou[1], Lina Yang[1], Tao Wang[1], Yu Wang[1], Zengqian Ding[1], Yanfu Wu[1], Kanglin Xiong[*1,2,a)], Jiagui Feng[*1,2,b)]

[1]Gusu Laboratory of Materials, Suzhou 215123, China

[2]Suzhou Institute of Nano-Tech and Nano-Bionics, CAS, Suzhou 215123, China

[a)] email: klxiong2008@sinano.ac.cn

[b)] email: jgfeng2017@sinano.ac.cn

Realization of practical superconducting quantum computing requires many qubits of long coherence time. Compared to the commonly used Ta deposited on c-plane sapphire, which occasionally form α-Ta (111) grains and β-tantalum grains, high quality Ta (110) film can grow epitaxial on a-plane sapphire because of the atomic relationships at the interface. Well-ordered α-Ta (110) film on wafer-scale a-plane sapphire has been prepared. The film exhibits high residual resistance ratio. Transmon qubits fabricated using these film shows relaxation times exceeding 150 μs. The results suggest Ta film on a-plane sapphire is a promising choice for long coherence time qubit on wafer scale.

Two ambitious goals being pursued in superconducting quantum computing are the noisy intermediate-scale quantum computing and fault-tolerant quantum computing, both of which require further scaling up the quantum chips with high coherence times of qubits across a large scale.[1-4] Significant enhancements in coherence times have been achieved by minimizing the sensitivity to decoherence sources through the design of qubit circuits.[5-9] On the other hand, eliminating decoherence channels such as two-level defects, which are closely related to the materials and device fabrication processes, paves the way for future steps.[10-15] To that purpose, novel materials[16-20] and innovative manufacturing technologies[21-24] are explored to improve the coherence times of superconducting qubits.

Recently, α-Ta (110) film has been used as a new material platform for building superconducting quantum circuits, boosting the coherence time of Transmon qubits to 300 μs;[18] shortly, this value was optimized to 500 μs.[19] Based on α-Ta (110) film, superconducting quantum chip with more than 100 qubits was also developed and the median coherence time exceeded 100 μs.[25] Studies explain this significant improvement of the coherence times using α-Ta (110) film can be attributed to the dense amorphous $Ta_2O_5$ passivation layer of low microwave loss formed on the surface of α-Ta (110) during piranha solution treatment.[26-28] This $Ta_2O_5$ is very stable and has little aging effect on the intrinsic quality factor of the superconducting resonator in the atmosphere for a long time.[28] However, the commonly used α-Ta (110) film grown on c-plane sapphire can occasionally form other structures, such as α-Ta (111) grains[26-27] and β-tantalum grains[29-31], which will affect the quality of the $Ta_2O_5$ layer[26] and impact the performance of superconducting qubit circuits. This is harmful to wafer scale high quality superconducting circuit fabrication. In contrast to α-Ta (110) film deposited on c-plane sapphire substrate, quasi-single crystal α-Ta (110) film can be epitaxial grown on wafer-scale a-plane sapphire substrate because of the atomic relationships at their interface.[32]

In this report, we present the sputtering deposition of a wafer scale well-ordered quasi-single crystal α-Ta (110) film with a low density of defects on a-plane sapphire. The α-Ta (110) film with thickness of 200 nm exhibits root mean square (RMS) below 0.7 nm over a 10 μm × 10 μm area. The $T_c$ value of the α-Ta (110) film is comparable to the bulk material, and the residual resistance ratio (RRR) is as high as 15.5. We also fabricate Transmon qubits using this α-Ta (110) film combining with Al/$AlO_x$/Al Josephson junctions. A wet etching process for Ta film is applied and optimized to achieve clean, smooth, and steep sidewalls. The observed relaxation time $T_1$ of qubits exceeds 150 μs. These results suggest that α-Ta (110) film grown on a-plane sapphire is a promising choice for wafer scale integration of superconducting qubits with long coherence times.

The atomic relationship at the interface between sapphire and α-Ta, which is crucial for determining the properties of the grown films. Perpendicular to the [11-20] direction, the $Al_2O_3$ lattice is composed of alternating layers of Al and O. The Al layer itself can be described as flat, with every third atomic position along the [0001] direction being vacant, as shown in Fig. 1(a). The α-Ta metal crystallizes in the body-centered cubic structure and its atomic structure on the (110) surface is shown in Fig. 1(b). At the interface between the (110) plane of the α-Ta and the (11-20) plane of sapphire, the two crystal structures align with each other. Specifically, their three-fold axes are oriented along the Ta-[1-11]//sapphire-[0001] direction. The spacing between the rows of atoms along the sapphire-[0001] and Ta-[1-11] directions is well matched, although there may be a small amount of mismatch. The rows of Al atom along [-2201] are closely aligned with the [001] rows of Ta atoms. These factors combine to form a convincing picture of the epitaxial formation of untwined Ta-(110) on the (11-20) plane, i.e., a-plane of sapphire. Additionally, there is a close match in the spacing between the Al and Ta planes normal to the growth direction, which allows for easy accommodation of surface steps on the sapphire substrate via metal overgrowth [Fig. 1(c)].

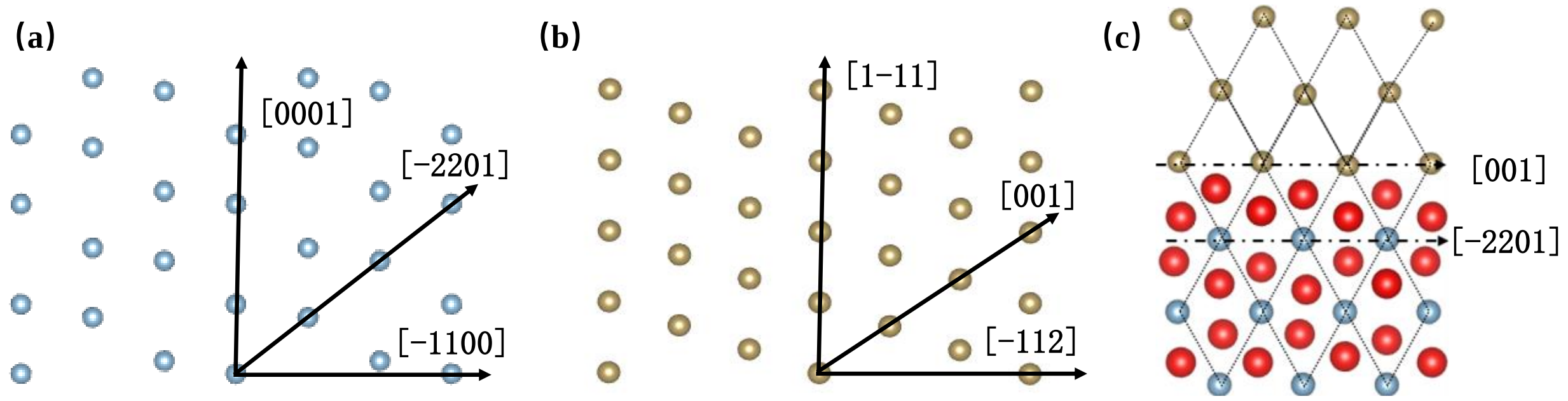

FIG. 1. Atomic arrangements. (a) Al sub-lattice in the a-plane sapphire. (b) Ta lattice on the (110) surface of α-Ta. (c) The interface atomic arrangement between a-plane sapphire and α-Ta (110) surface. The light blue spheres represent aluminum atoms, the golden spheres represent tantalum atoms, and the red spheres represent oxygen atoms.

Ta films with a thickness of 200 nm were deposited on 2-inch a-plane sapphire substrates using a DC-magnetron sputtering system. The a-plane sapphire substrate underwent wet-chemical cleaning, and then, was transferred to the UHV sputtering chamber. Prior to the film deposition, the substrate was thermally cleaned inside the sputtering chamber at 850 ℃ for 30 minutes, and then slowly cooled down to 700 ℃ at a rate of 30 ℃ per minute. During the deposition process, the substrate was maintained at 700 ℃. Ar gas at a rate of 28 sccm was continuously flowed into the chamber. The sputtering pressure was kept constant at 10 mTorr, and the DC sputtering power was set to 200 W. The deposited Ta films were characterized using X-Ray Diffraction (XRD), Atomic Force Microscope (AFM), and High-Angle Annular Dark-Field Scanning Transmission Electron Microscopy (HAADF-STEM). The temperature dependence of Ta film resistance was measured using a four-point method over a square area of 10 mm $\times$ 10 mm.

Fig. 2(a) shows the XRD result of the deposited Ta films. It can be clearly seen that the dominative features of the Ta films are the (110) and (220) diffraction peaks at 38.47$^{o}$ and 82.46$^{o}$ respectively, which are consistent with the lattice constant of bulk α-Ta with bcc crystal structure. Apart from diffraction peaks of α-Ta (110), neither α-Ta with other crystal orientations nor β-Ta is observed in the full range of 2θ, suggesting the formation of quasi-single crystal α-Ta (110) films. The AFM image measured over a 10 μm × 10 μm area of the Ta films is shown in Fig. 2(b), resulting a surface RMS below 0.7 nm. The inset of Fig. 2(b) shows the AFM image over a 2 μm × 2 μm area. A typical rectangular-shape feature of the grains is evident, and the grains look highly homogeneous in size and are uniformly distributed on the surface. The rectangular grains orient along the same direction, with rare occurrence of twinned crystals. These AFM features altogether provide evidence convincing the growth of well-ordered quasi-single crystal α-Ta (110) films with a low density of defects. The close microstructure analysis near the hetero-interface between α-Ta film and sapphire by high-resolution HAADF-STEM is shown in Fig. 2(c), which illustrates that a sharp interface with no intermixing is formed between the α-Ta (110) film and sapphire substrate. The atomic arrangement of Ta resembles that of Al in the substrate [inset of Fig. 2(c)], which is consistent with what is anticipated and schematically illustrated in Fig. 1(c). Moreover, the characterization by HAADF-STEM reveals the formation of a dense and uniform $Ta_2O_5$ amorphous oxide layer on the film surface, with an average thickness of about 2.76 nm (Fig. S1 in the Supplementary Material).

The temperature dependence of the Ta film resistance is shown in Fig. 2(d). From the curve, it can be seen the superconducting transition temperature of the film is about 4.33K, with a transition temperature range of 0.1K [inset of Fig. 2(d)]. We calculate RRR as a ratio of measured resistance between 300 K and 10 K, and the value is as high as 15.5. This value is more than 3 times larger than that (4.5) of the Ta film grown on c-plane sapphire, with which the qubit with a coherence time of 500 μs was realized.[19] RRR has long been used to determine the quality of superconducting films for constructing superconducting radiofrequency cavities.[33-34] Recent study shows that RRR can also serve as a gauge for superconducting qubit performance.[35] Thus, the high RRR value suggests qubits with long coherence times can be achieved by using this kind of α-Ta (110) film as the superconducting circuit material.

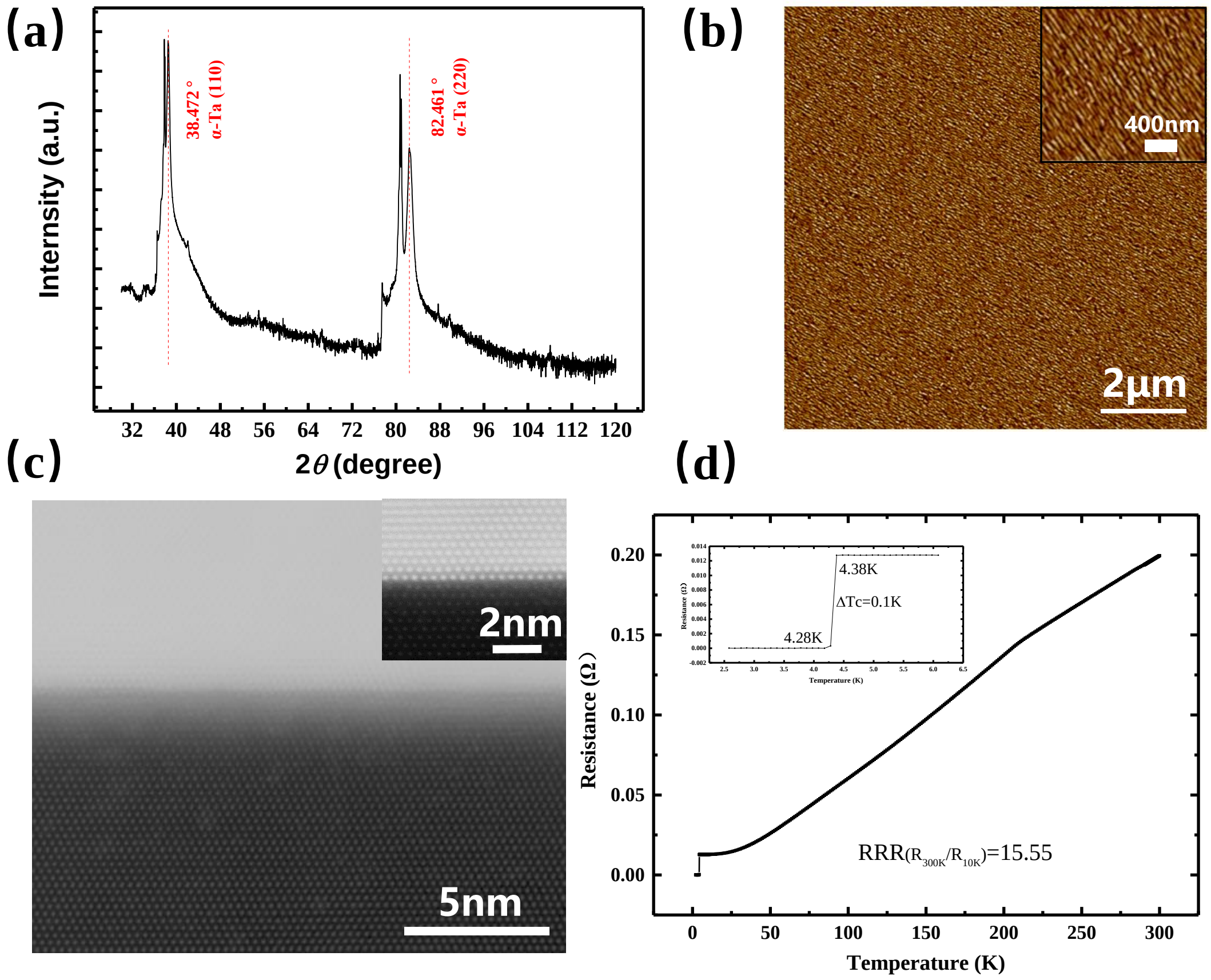

FIG. 2. Structure and performance characterization of α-Ta (110) film. (a) XRD characterization of the film with red dashed lines indicating the (110) and (220) peak positions. (b) AFM image of surface morphology, the inset is a zoom-in image. (c) STEM characterization of the interface between the film and substrate. The inset is a zoom-in image, which indicates that the atomic arrangement of Ta resembles that of Al in the substrate. (d) Measurement curve of film resistance vs temperature, with inset showing the superconducting transition temperature.

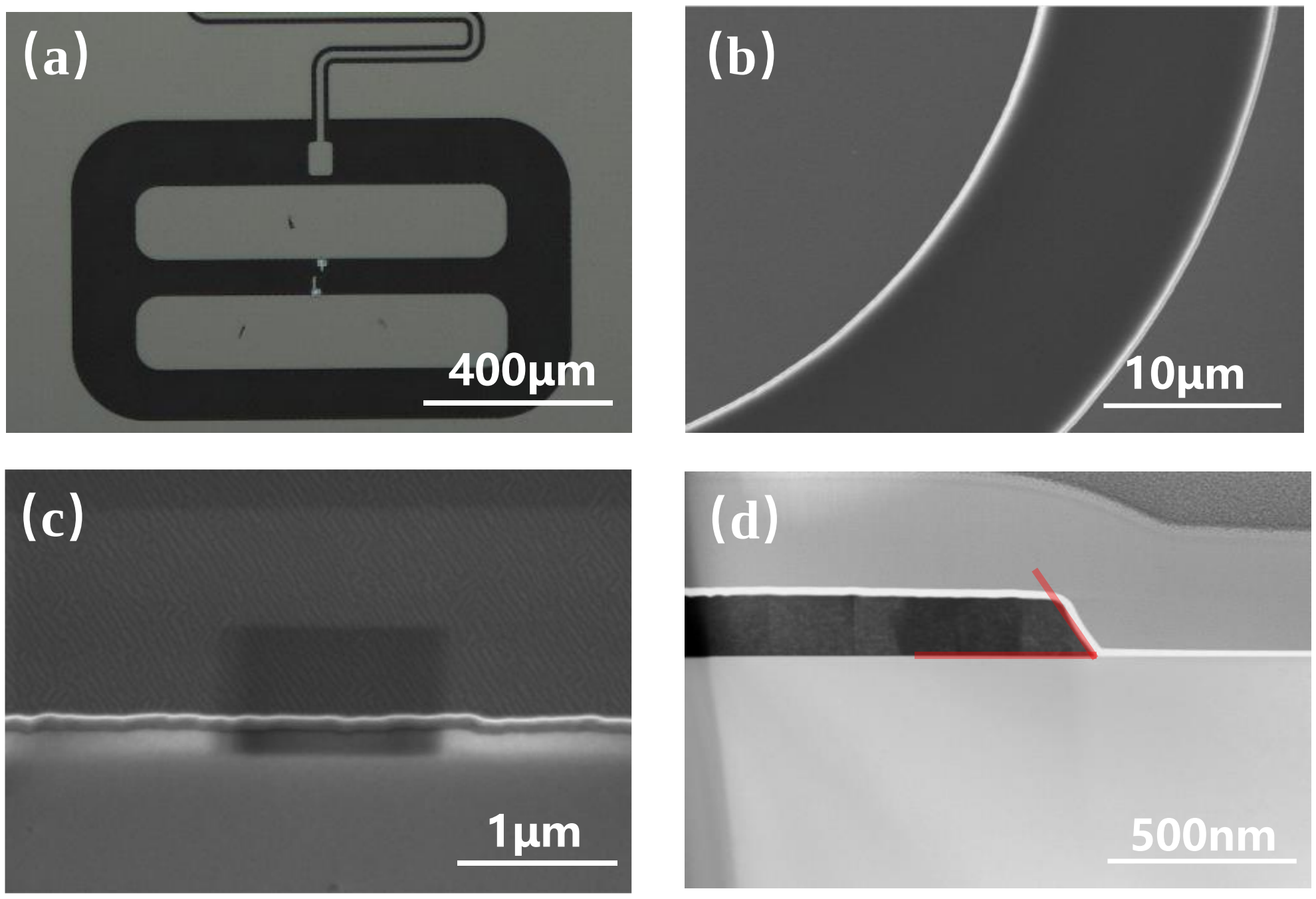

FIG. 3. (a) Optical microscope image of a representative Transmon qubit. (b) and (c) SEM images of the wet-etched α-Ta (110) film showing the smoothness and cleanness of the sidewalls and nearby α-Ta (110). (d) STEM image of the interface between the etched α-Ta (110) film and substrate, where the light red angle indicates the sidewall-etching angle.

Transmon qubits with geometry shown in Fig. 3(a) are fabricated as follows, using the α-Ta (110) film deposited on a-plane sapphire. The sapphire wafers with 200 nm Ta films are sonicated in acetone and isopropyl alcohol for 5 min each, and rinsed by DI water, then cleaned in a piranha solution for 20 min to form a dense, uniform, stable, and reliable $Ta_2O_5$ passivation layer with low loss. Patterns of the capacitors, resonators and drive lines are defined by photolithography with a layer of resist of AZ1500 series. Wet etching method is applied to remove the unwanted regions of Ta film. A mixture of HF, $HNO_3$ and $H_2O$ solution is used to etch the Ta film. To strip resist, devices are dipped in NMP for 2.5 h at 80 ℃. The mixture ratio of the HF, $HNO_3$ and $H_2O$ solution and etching time are optimized to achieve clean, smooth, and steep sidewalls, which likely help minimize the microwave loss volume of the sidewalls in superconducting circuits and improve the relaxation time $T_1$.[18] As shown in Fig. 3(b) and (c), the pattern edges and nearby α-Ta (110) film are clean and smooth with roughness in sub-micrometer scale. Fig. 3(d) shows the morphology of the sidewalls. The etching angle to the substrate is steep, and there is no visible residue at edges.

The next step is Al/AlOx/Al Josephson junction fabrication using Dolan bridge technique. Electron-beam lithography is used to pattern the junctions. Before the double angle evaporation, Ar ion milling is used to clean the Ta film surface to achieve superconducting connection at the contact regions. We then deposit 30 nm Al for the lower layer and 60 nm Al for the top layer of junctions, with an oxidization between two depositions.

TABLE I. Device parameters of six qubits. $\omega_r$ and $\omega_{01}$ are the readout cavity frequency and qubit frequency.

| | qubit1 | qubit2 | qubit3 | qubit4 | qubit5 | qubit6 |
|---|---|---|---|---|---|---|
| $\omega_r/2\pi$ (GHz) | 6.591 | 6.586 | 6.787 | 6.788 | 7.302 | 7.3 |
| $\omega_{01}/2\pi$ (GHz) | 4.058 | 3.848 | 4.207 | 4.4626 | 3.97 | 3.933 |
| $T_1$ (μs) | 150 | 112.5 | 106.2 | 80.6 | 97 | 89 |
| $T_2$ (μs) | 22.1 | 27.2 | 25.7 | 30.7 | 22 | 38.6 |

The relaxation time $T_1$ and coherence time $T_2$ of six qubits fabricated on the α-Ta (110) film deposited on a-plane sapphire wafer with the same processes were characterized. Devices were wire-bonded in Al sample boxes, loaded into dilution fridge, and cooled down to ~10 mK. The relaxation time $T_1$ is measured by applying a π-pulse signal to prepare the qubit in the excited state and reading the qubit state after a delay time. For each $T_1$ measurement, 1500 repetitions are used to obtain the mean decay plot. A typical $T_1$ measurement plot is shown in Fig. 4(a), the qubit excited state population $P_1$ is recorded as a function of delay time t, and the relaxation time $T_1$ is acquired by fitting the result to an exponential decay function. The coherence time $T_2$ is obtained from an Echo sequence, which involves two π/2-pulses separated by a variable time delay with a π-pulse to invert the qubit state in between. The resulted coherence properties for all six qubits are shown in Table I, with $T_1$ ranging from 80 μs to 150 μs, and $T_2$ ranging from 22 μs to 39 μs. The difference in $T_1$ and $T_2$ could be from the variations of fabrication, impurities, or surface roughness. The values of $T_2$ for all six qubits are close and relatively low compared to the $T_1$, which are possibly limited by the magnetic fluctuations or temperature fluctuations of the measurement system. $T_1$ of the best device qubit1 was measured repeatedly for a long-time interval, as shown in Fig. 4(b). The long-time averaged $T_1$ of qubit1 is around 150 μs, and the highest $T_1$ trace is plot in Fig. 4(a), which yields a value near 198 μs. The $T_1$ values of the devices are expected to be further improved through the optimization of fabrication, packaging, and measurement system.

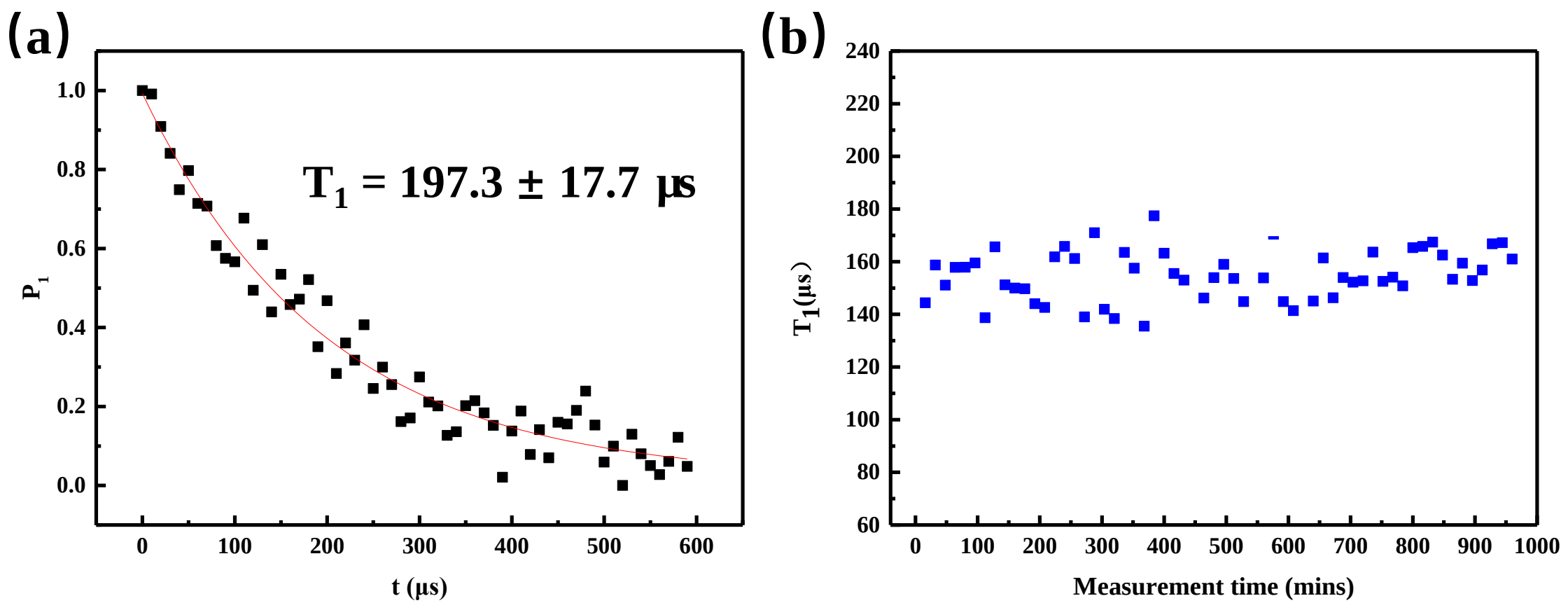

FIG. 4. The characterization of relaxation times for superconducting qubits. (a) A typical decay curve of the qubit with $T_1$ approximately 198μs. (b) Long-time repeated measurements of $T_1$ for qubit1.

In summary, using a-plane sapphire substrate with a close lattice match to α-Ta, wafer-scale well-ordered quasi-single crystal α-Ta (110) films with low defects density are sputtering deposited. The resulting α-Ta (110) films, which are 200 nm thick, exhibit an RMS below 0.7 nm over an area of 10 μm × 10 μm. The $T_c$ values of the α-Ta (110) films are consistent with the bulk material of 4.2 K, and the RRR is as high as 15.5. High-performance superconducting qubits using these α-Ta (110) thin films are also fabricated with the application of a wet etching process for Ta films. The observed $T_1$ of qubits exceeds 150 μs, and the highest $T_1$ is about 198 μs. These findings suggest that α-Ta film grown on a-plane sapphire could be a promising choice for wafer scale integration of qubits with long coherence times.

**Acknowledgements**

K. L. X acknowledges support from the Youth Innovation Promotion Association of Chinese Academy of Sciences (2019319). J. G. F. acknowledges support from the Start-up foundation of Suzhou Institute of Nano-Tech and Nano-Bionics, CAS, Suzhou (Y9AAD110).

**Data availability statement**

The data that support the findings of this study are available from the corresponding author upon reasonable request.